\documentclass[a4paper]{revtex4}
\usepackage[dvipdfmx]{graphicx}
\usepackage{fancyhdr}
\usepackage{amsmath}
\usepackage{color}

\begin{document}

\title{Scale invariant extension of the standard model with a hidden QCD-like sector}

%

\author{H. Hatanaka}
\affiliation{School of Physics, Korea Institute for Advanced Study, Seoul 130-722, 
Republic of Korea}

\begin{abstract}
A scale-invariant extension of the Standard Model with a singlet-scalar and hidden-QCD sector is studied. The electroweak symmetry breaking scale is generated dynamically by an
asymptotic-free hidden-QCD sector, and mediated by the Higgs-singlet coupling.
Hidden-QCD pions are stable and can be a candidate of the cold dark matter. 
This presentation is based on the collaboration with D.~W.~Jung and P.~Ko \cite{HJK}.
\end{abstract}

\maketitle

\thispagestyle{fancy}


\section{Introduction}

Even after the discovery of the Higgs particle at the LHC, the origin of the electroweak scale is still an unsolved problem.
Here we consider a scale-invariant extension of the Standard Model with a singlet scalar and hidden non-Abelian gauge theory\cite{HK}.

The model contains the SM fields, a singlet scalar $S$ and 
a scale-invariant hidden QCD sector. The Lagrangian is given by
\begin{eqnarray}
{\cal L} 
&=&
{\cal L}_{\rm SM} (\mu_H^2 = 0) 
+ \frac{1}{2} (\partial_\mu S)^2 
 -  \frac{\lambda_S}{8} S^4
 + \frac{\lambda_{HS}}{2} H^\dag H S^2 
\nonumber \\
&&
 - \frac{1}{2} \text{tr}\, {G}_{\mu\nu} G^{\mu\nu} 
+ \sum_{k=1}^{N_{h,f}} \bar{Q}_k (i\gamma^\mu D_\mu - \lambda_{Q,k} S) Q_k,  
\end{eqnarray}
where $\mu_H$ is the mass parameter of the SM Higgs. $G_{\mu\nu}$
is the field strength of the hidden-QCD with $SU(N_{h,c})$ gauge symmetry.
The SM singlet scalar $S$ couples to the hidden-QCD quarks $Q_f$ through the Yukawa interaction $\lambda_Q$.
Since there are no dimension-full parameters in the Lagrangian, this system is scale-invariant at a classical level.
Thanks to the asymptotic-free behavior of the hidden-QCD,
at a quantum level and at a low-energy scale, the hidden-QCD quarks can condensate
$\langle \bar{Q}Q \rangle$, which induce a linear term of $S$.
Then the potential of $S$ can be tilted and $S$ can develop a vacuum expectation value (VEV).
The VEV of the singlet scalar generates a Higgs mass term $-\frac{\lambda_{HS}}{2} \langle S\rangle^2 H^\dag H$. 
We assume that $\lambda_{HS} > 0$ so that non-zero $\langle S \rangle$ triggers the electroweak symmetry breaking.

\section{Linear sigma model}

Hereafter we consider the case in which $N_{h,c}=3$, $N_{h,f}=2$
and 
$\lambda_Q = \text{diag}(\lambda_{Qu},\lambda_{Qd})$,
$\lambda_{Qu} \sim \lambda_{Qd}$.
Then the low-energy effective theory of the hidden-QCD is described by the pi-meson triplets and the sigma meson. 
It can be written in the form of a linear-sigma model
\begin{eqnarray}
{\cal L}_{L\sigma M}
&=& \frac{1}{2} (\partial_\mu \Sigma)^2  + \frac{1}{2} (\partial_\mu \pi)^2
 - \frac{\lambda_{\sigma}}{4} (\Sigma^2  + \pi^2)^2
+ \frac{\mu_\sigma^2}{2} (\Sigma^2  + \pi^2)
 + m_{S\sigma}^2 S\Sigma,
\end{eqnarray}
where $\Sigma$ and $\pi$ represents sigma and pi meson fields, respectively.
We parameterize the vacuum expectation values and fluctuations of scalars as
\begin{eqnarray}
H = \begin{pmatrix}0, & v_H + h \end{pmatrix}^T/\sqrt{2},
\quad
S = v_S + s,
\quad
\Sigma = v_\sigma + \sigma.
\end{eqnarray}
Three vacuum conditions reduce the number of free parameters. Furthermore, two parameters $\lambda_\sigma$, $\mu_\sigma$ are traded with the pion mass $M_\pi$ and a sigma meson mass parameter $M_{\sigma\sigma}$.
Hence the scalar mass matrix ${\cal L} \supset -\frac{1}{2} (h,s,\sigma){\cal M}(h,s,\sigma)^T$ 
takes the form of
\begin{eqnarray}
{\cal M} &=& 
\begin{pmatrix} 
M_{hh}^2 & M_{hs}^2 & 0 \\
M_{hs}^2 & M_{ss}^2 & -m_{S\sigma}^2 \\
0 & -m_{S\sigma}^2 & M_{\sigma\sigma}^2
\end{pmatrix},
\label{eq:mixmass}
\end{eqnarray}
with
\begin{align}
M_{hh}^2 &= \lambda_H v_H^2 = \lambda_{HS}v_H^2 \tan^2\beta,
&M_{hs}^2 &= -\lambda_{HS} v_H^2 \tan\beta,
&M_{ss}^2 &= \lambda_{HS} v_H^2 \left( 1 + \frac{3 M_\pi^2 F_\pi^2}{\lambda_{HS} \tan^2\beta v_H^4}\right),
\\
-m_{S\sigma}^2 &= -\frac{M_\pi^2 F_\pi}{v_S},\label{eq:mixing}
\end{align}
where $\tan\beta \equiv v_S/v_H$,
$v_H = 246\text{GeV}$
and
$F_\pi \equiv v_\sigma$.
Since one of the physical scalar should be the Higgs boson with mass $m_H^2 = 
(125\text{GeV})^2$ is  one of the eigenvalues of ${\cal M}$,
one can use a condition $\det({\cal M} - m_H^2 I_3) = 0$ ($I_3$ is a $3\times3$ unit matrix) to express $\lambda_{HS}$ in an analytical form.
We parameterize $M_{\sigma\sigma} = \xi_\sigma F_\pi$.

At this stage we have four free parameters: $v_S$, $v_\sigma\equiv F_\pi$, $M_\pi$ and $M_{\sigma\sigma}$ (or $\xi_\sigma$).
To reduce the number of free parameters, in particlar, to relate the $M_{\sigma\sigma}$ with $F_\pi$, we use a holographic treatment of the QCD.
We follow mainly \cite{RP1,RP2}, but unlike the original work by Rold and Pomarol\cite{RP2} in which the lightest scalar meson resonance is identified with $a_0(980)$,
we regard the lightest scalar meson state as the sigma meson.
and we imposed a new condition for sigma-singlet mixing
\begin{eqnarray}
m_{S\sigma}^2 = F_\sigma M_\sigma
\end{eqnarray}
where $F_\sigma$ and $M_\sigma$ are the decay constant and mass of the sigma meson,respectively. Using \eqref{eq:mixing} and Gell-Mann--Oaks-Renner relation the above relation is written by
\begin{eqnarray}
F_{\sigma} M_{\sigma} = B_0 F_\pi,
\quad
B_0 \equiv \langle \bar{Q} Q \rangle/F_\pi^2,
\end{eqnarray}

\begin{figure}[ht]
\includegraphics[width=0.5\linewidth]{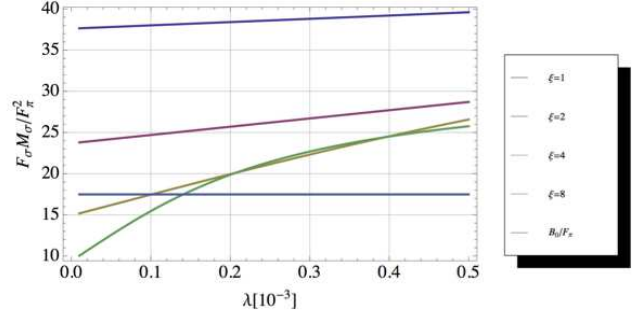}
\caption{Value of $F_\sigma M_\sigma$ for $\sigma=S^{(1)}$ in the AdS/QCD\cite{RP2}
in the unit of $F_\pi^2$.}\label{fig:mixingcond}
\end{figure}
\begin{figure}[ht]
\includegraphics[width=0.5\linewidth]{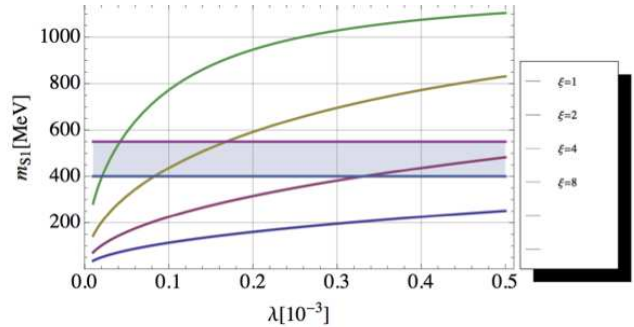}
\caption{Mass of the lightest scalar meson calculated in AdS/QCD framework\cite{RP2}. We regard this scalar as sigma meson. Gray band indicates the allowed mass range of $f_0(500)$. We also plot $B_0 F_\pi = 17.5 F_\pi^2$ as the horizontal solid line.}\label{fig:sigma}
\end{figure}

For $\xi=4$ we obtain for $\lambda \simeq 1.0 \times 10^{-4}$.
We also find that with this value we obtain $M_\sigma \simeq 5 F_\pi$ ($\xi_\sigma \simeq 5$).
This value well agree with the observed $f_0(500)$ meson mass $m_{f_0(500)} = 400-550\text{MeV}$ \cite{PDG} (Fig. \ref{fig:sigma}).

\section{Numerical Analysis (Preliminary)}

In this preliminary study, we fix $F_\pi = 1000\text{GeV}$ and consider
a parameter region
$50 \text{GeV} \le  M_\pi \le 300 $ and $0 \le \tan\beta \le 16$.

Three scalar particles $(h,s,\sigma)$ is mixed with each other by the mass matrix \eqref{eq:mixmass}.
Mass-eigenstates are $h',h_1,\sigma$:
\begin{eqnarray}
\begin{pmatrix} h \\ s \\ \sigma \end{pmatrix}
=
\begin{pmatrix} 
\rho_{hh'} & \rho_{hh_1}& \rho_{h\sigma'} \\
\rho_{sh'} & \rho_{sh_1}& \rho_{s\sigma'} \\
\rho_{\sigma h'} & \rho_{\sigma h_1}& \rho_{\sigma\sigma'} 
\end{pmatrix}
\begin{pmatrix} h' \\ h_1 \\ \sigma' \end{pmatrix},
\end{eqnarray}
where $h'$ corresponds to the physical Higgs boson and $m_{h'} = 125\text{GeV}$.
$m_{\sigma'} \sim M_{\sigma\sigma} \simeq 5F_\pi$.
$h_1$ can be heavier or lighter than the physical Higgs $h'$.
The signal strength of the Higgs production and decay will be suppressed by a mixing matrix element $\rho_{hh'}$:
\begin{eqnarray}
\frac{\mu_{\rm SIM}(\bar{p}p \to h\to \bar{f}f)}{\mu_{\rm SM}(\bar{p}p \to h\to \bar{f}f)} 
\simeq |\rho_{hh'}|^2 .
\end{eqnarray}
For the signal strength of our scale invariant model $\mu_{\rm SIM}$, we imposed a constraint
\begin{eqnarray}
|\rho_{hh'}|^2 &\ge& 0.9.  \label{mix-bound}
\end{eqnarray}
in the numerical study.

Since the hidden-pion is stable and interacts with SM particles weakly,
this particle can be the dark matter.
We calculate the relic density and spin-independent DM-nucleon scattering cross section 
of the hidden pion. In the numerical study, we used micrOMEGAs.
Because the lightest baryonic states can also be a stable dark matter, we have 
imposed only an upper bound of 2-sigma upper limit of the observed relic density
\begin{eqnarray}
\Omega_{{\rm DM}=\pi_h} h^2 \lesssim 0.14.
\label{DM-bound}
\end{eqnarray}

\begin{figure}[ht]
\includegraphics[width=0.4\linewidth]{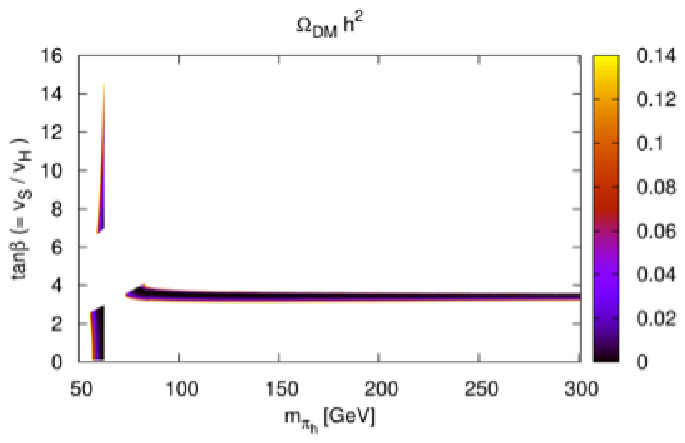}
\includegraphics[width=0.4\linewidth]{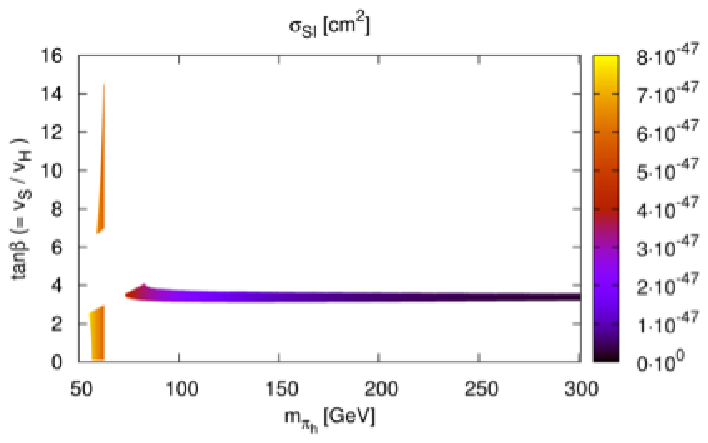}
\caption{Relic density and DM-nucleon cross section.
[Left] Relic density of the hidden pion as cold dark matter
[Right] the spin-independent DM-nucleon cross sections
of the hidden-pion as cold dark matter.}\label{fig:relic}
\end{figure}

In Fig. \ref{fig:relic}, we plotted the allowed region of the model.
The colored region is allowed by both \eqref{mix-bound} and \eqref{DM-bound}.
The allowed parameter space consists of three disconnected regions:
(a) $m_{\pi_h} \sim 60\text{GeV}$ and $\tan\beta \gtrsim 6$
(b) $m_{\pi_h} \sim 60\text{GeV}$ and $\tan\beta \lesssim 4$
(c) $\tan\beta \sim4$ and $m_{\pi_h} \gtrsim 70\text{GeV}$.
A parameter region $m_{\pi_h} \sim 70\text{GeV}$ and $\tan\beta \sim5$
is forbidden since \eqref{mix-bound} cannot be satisfied.

We also plotted the spin-independent nucleon-DM cross section (Fig. \ref{fig:relic}, right).
We found that the cross section can be small sufficiently and can easily evade the experimental limit of direct detection experiments\cite{LUX}.

\begin{figure}[ht]
\includegraphics[width=0.4\linewidth]{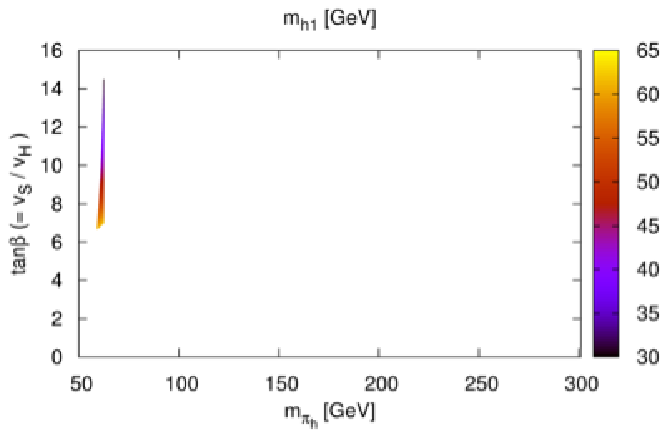}
\includegraphics[width=0.4\linewidth]{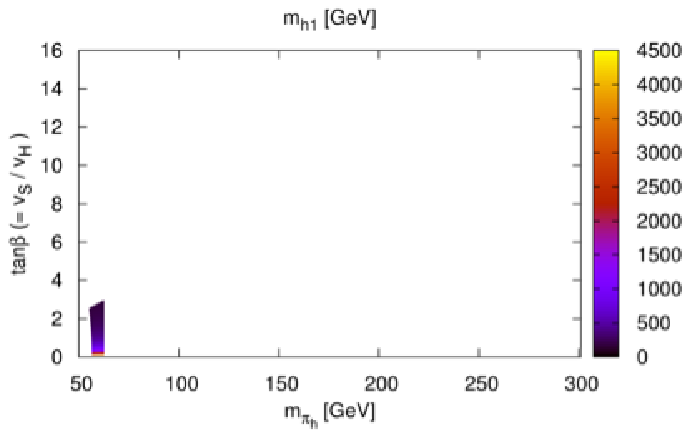}
\includegraphics[width=0.4\linewidth]{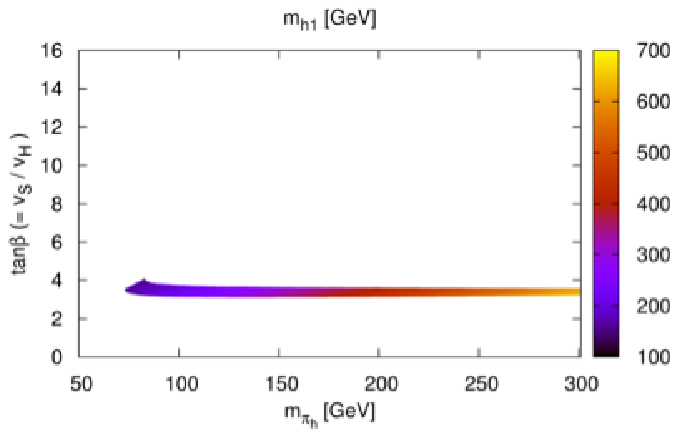}
\caption{Masses of the scalar $h_1$ ($m_{h_1} \neq 125\text{GeV}$) for three allowed regions.
(upper left) $m_\pi \sim m_{H}/2$ and $\tan\beta \gtrsim7$.
(lower) $m_\pi \sim m_{H}/2$ and $\tan\beta \lesssim 4$, 
(upper right) $m_{h_1} \simeq 2 m_{\pi_h}$.
}\label{fig:masses}
\end{figure}

We have also measured the mass of the non-SM scalar particle $h_1$
in Fig.\ref{fig:masses}.
In regions (a) and (b) the annihilation cross section of the DM is enhanced through the $m_{H}$ resonance since $m_{H} \sim 2 m_{\pi_h}$,
whereas in (c) annihilation cross section is raised due to the $h_1$ resonance.

\section{Summary}
A scale-invariant extension of the SM with a singlet scalar and hidden-QCD sector is studied. We reformulated the hidden-QCD sector in terms of the linear-sigma model.
A further study will be presented in an another paper\cite{HJK}.

%





\begin{acknowledgments}
This work is partly supported by 
NRF Research Grant 2012R1A2A1A01006053 of the Republic of Korea.
\end{acknowledgments}

\bigskip 

\begin{thebibliography}{99} 
\bibitem{HJK} 
  H.~Hatanaka, D.~W.~Jung and P.~Ko,
  JHEP08(2016)094
  [arXiv:1606.02969 [hep-ph]].

\bibitem{HK} 
  T.~Hur and P.~Ko,
  Phys.\ Rev.\ Lett.\  {\bf 106}, 141802 (2011)
  [arXiv:1103.2571 [hep-ph]].


\bibitem{RP1} 
  L.~Da Rold and A.~Pomarol,
  Nucl.\ Phys.\ B {\bf 721}, 79 (2005)
  [hep-ph/0501218].

\bibitem{RP2} 
  L.~Da Rold and A.~Pomarol,
  JHEP {\bf 0601}, 157 (2006)
  [hep-ph/0510268].

\bibitem{PDG} 
  K.~A.~Olive {\it et al.}  [Particle Data Group Collaboration],
  Chin.\ Phys.\ C {\bf 38}, 090001 (2014).

\bibitem{LUX} 
  D.~S.~Akerib {\it et al.}  [LUX Collaboration],
  Phys.\ Rev.\ Lett.\  {\bf 112}, 091303 (2014)
  [arXiv:1310.8214 [astro-ph.CO]].

\end{thebibliography}

\end{document}